# Nanotechnology as a Field of Science:

# Its Delineation in terms of Journals and Patents

*Revision, 24 September 2006*

Loet Leydesdorff [a] & Ping Zhou [b]

[a] Amsterdam School of Communications Research (ASCoR),

University of Amsterdam, Kloveniersburgwal 48, 1012 CX Amsterdam, The Netherlands

loet@leydesdorff.net; http://www.leydesdorff.net

[b] Institute of Scientific and Technical Information of China,

15 Fuxing Road, Beijing, 100038, P. R. China;

zhoup@istic.ac.cn; pzhou@fmg.uva.nl.

**Abstract**

The Journal Citation Reports of the *Science Citation Index* 2004 were used to delineate a core set of nanotechnology journals and a nanotechnology-relevant set. In comparison with 2003, the core set has grown and the relevant set has decreased. This suggests a higher degree of codification in the field of nanotechnology: the field has become more focused in terms of citation practices. Using the citing patterns among journals at the aggregate level, a core group of ten nanotechnology journals in the vector space can be delineated on the criterion of betweenness centrality. National contributions to this core group of journals are evaluated for the years 2003, 2004, and 2005. Additionally, the specific class of nanotechnology patents in the database of the U.S. Patent and Trade Office (USPTO) is analyzed to determine if non-patent literature references can be used as a source for the delineation of the knowledge base in terms of scientific journals. The references are primarily to general science journals and letters, and therefore not specific enough for the purpose of delineating a journal set.

## 1. Introduction

The emergence of new fields of science and technology potentially upsets previously existing classification systems. Chan (1999, pp. 12-16) explains that the Library of Congress of the United States (LC), for example, is based on “literary warrant.” A classification scheme based on literary warrant is not logically deduced from some abstract philosophical system for classifying knowledge but inductively developed in reference to the holdings of a particular library, or to what is or has been published. In other words, it is based on what the actual literature of the time warrants. The LC has a policy of continuous revision to take current literary warrant into account, so that new areas are developed and obsolete elements are removed or revised (Leydesdorff & Bensman, 2006).

Similarly, the U.S. Patent and Trade Office (USPTO) decided in 2004 to introduce a new category into its classification scheme devoted to "nanotechnology." As defined by the USPTO (at http://www.uspto.gov/web/patents/biochempharm/crossref.htm), nanotechnology patents in this "Class 977" must meet the following criteria:

- Relate to research and technology development in the length scale of approximately 1-100 nm in at least one dimension;
- Provide a fundamental understanding of phenomena and materials at the nanoscale, and create and use structures, devices, and systems that have size-dependent novel properties and functions.

Patents issued before the new class was created have actively been reclassified by the office. In the meantime, the sub-classifications of Class 977 contain more than 250 categories.[1]

In summary, these two catalogues are very detailed, but they potentially suffer from so-called indexer effects (Courtial *et al*., 1984, 1993; Healey *et al*., 1986; King, 1987; Leydesdorff, 1989). Indexes can be considered as second-order mechanisms of codification, while publication and citation practices by active scientists provide first-order updates of scientific literature (Leydesdorff, 2002). In the case of patent references, examiners add citations to the references provided in the applications, but one may expect this to be the case in terms of previous patents more than in the case of previous non-patent literature references (NPLR). NPLRs are less central to the legal upholding of a patent when litigated in court (Granstrand, 1999; Jaffe & Traitenberg, 2002; Meyer,

[1] A similar effort is ongoing at the European and Japanese Patent Offices (Sheu *et al*., 2006). We use the USPTO database in this study because it uses a mark-up language (html), while the EPO database is pdf-based and also otherwise less accessible for online investigations (Leydesdorff, 2004).

2000, forthcoming). Can one use aggregated citations among journals and/or in classes of patents for the delineation of a nano-relevant and core nano set of journals?

In this study, we update on a previous attempt (Zhou & Leydesdorff, 2006) to use the Journal Citation Reports of the *Science Citation Index* 2003 for the construction of a nano-relevant set of journals. We use the JCR-data of 2004,[2] and extend the previous analysis by using betweenness centrality as a measure of interdisciplinarity (Leydesdorff, 2006a). "Betweenness centrality" will be analyzed both in the set of journals cited by seed journals in nanoscience and nanotechnology and in the set which is citing this set. Finally, we use non-patent literature references (NPLR) in patent class 977 to examine whether and how a bridge with the relevant journal literature might be provided (Meyer & Persson, 1997).

## 2. Methods and materials

The aggregated journal-journal citation data was harvested from the Journal Citation Reports of the *Science Citation Index* 2004. This data was brought under the control of relational database management. This enables us to generate files that can be imported into programs for statistical analysis and visualization. We use SPSS, UCINet, and Pajek for the statistical analysis, and the latter program also for the visualization.

The data allows us to generate citation environments for individual source journals or for a list of such journals at a variable threshold level. (In most analyses below the threshold was one percent of the total citations in the respective dimension (He & Pao, 1986; Leydesdorff & Cozzens, 1993).) The data matrix of aggregated citations among journals is asymmetrical and therefore contains structures in both the "citing" and the "cited" directions. These structures are analyzed using factor analysis with Varimax rotation.

[2] At the time of this research (June 2005), CD-Rom versions of the JCR 2005 were not yet available.

Visualizations are based on the vector-space model, using the cosine between vectors as the similarity measure (Salton & McGill, 1983) and the spring-embedded algorithm of Kamada & Kawai (1989) for the representation. The visualizations correspond by and large with the results of factor analysis, since the Pearson correlation coefficient—which is basic to factor analysis—and the cosine are similar, except that the latter normalizes on the basis of the geometrical mean while the former uses the arithmetic mean (Jones & Furnas, 1987; Ahlgren *et al*., 2003; Chen, 2006).

The patent data was downloaded from the USPTO database on June 20, 2006, by using the Internet module available in Visual Basic (Leydesdorff, 2004, at p. 1001) and the search string "CCL/977/$ and ISD/$/$/2005".[3] The data was then brought under the control of a database manager for further processing. The descriptive statistics of this data are provided in Table 1.

[3] The number of patents in Class 977 of the USPTO is declining since 2003 (723 patents). It has been suggested that the Chinese have changed their policy of patenting in the U.S.A. (Caroline Wagner, *personal communication*). I have no explanation why the numbers are so much lower than those reported for the database of the European Patent Office by Sheu *et al*. (2006).

| **"CCL/977/$ and ISD/$/$/2005"** | |
|---|---|
| Number of patents retrieved | 336 |
| Nr of assignees | 352 |
| Nr of inventors | 1027 |
| Nr of patent references | 4830 |
| NPLR | 1948 |

**Table 1:** Patents assigned under the category "nanotechnology" in the USPTO database during 2005.

### 3. Nanotechnology journals

Zhou & Leydesdorff (2006) used three journals included in *Science Citation Index* 2003 with the stem "nano" in their title. In 2004, six such journals could be retrieved in the *Science Citation Index*.[4] Table 2 shows the aggregated citation matrix among these six journals.

| | *citing →* | | | | | |
|---|---|---|---|---|---|---|
| *Fullerenes Nanotubes and Carbon Nanostructures* | 21 | **0** | **0** | **0** | **0** | **0** |
| *IEEE Transaction on Nanotechnology* | **0** | 33 | 0 | 3 | 15 | 18 |
| *Journal of Nanoparticle Research* | **0** | 2 | 22 | 7 | 13 | 7 |
| *Journal of Nanoscience and Nanotechnology* | **2** | 2 | 0 | 30 | 10 | 9 |
| *Nano Letters* | **0** | 23 | 9 | 96 | 727 | 160 |
| *Nanotechnology* | **2** | 24 | 7 | 23 | 107 | 247 |

**Table 2**: Aggregated citation matrix among six journals with the stem "nano" in their main title and included in the *SCI* 2004.

This citation matrix reveals upon inspection that *Fullerenes Nanotubes and Carbon Nanostructures* is not cited by the other five journals, and articles in this journal rarely cite papers in the other ones. *Fullerenes Nanotubes and Carbon Nanostructures* is an older journal; the journal is a prolongation of *Fullerene Science and Technology*, which had published its first volume in 1992. (Nanotubes were discovered as a specific form of fullerenes in 1991.)

[4] Two more journals contain the stem "nano" in their subtitles: *Microsystems Technology: Micro- and Nano-Systems, Information Storage and Processing* and *Physica E: Low-dimensional Structures and Nanostructures.*

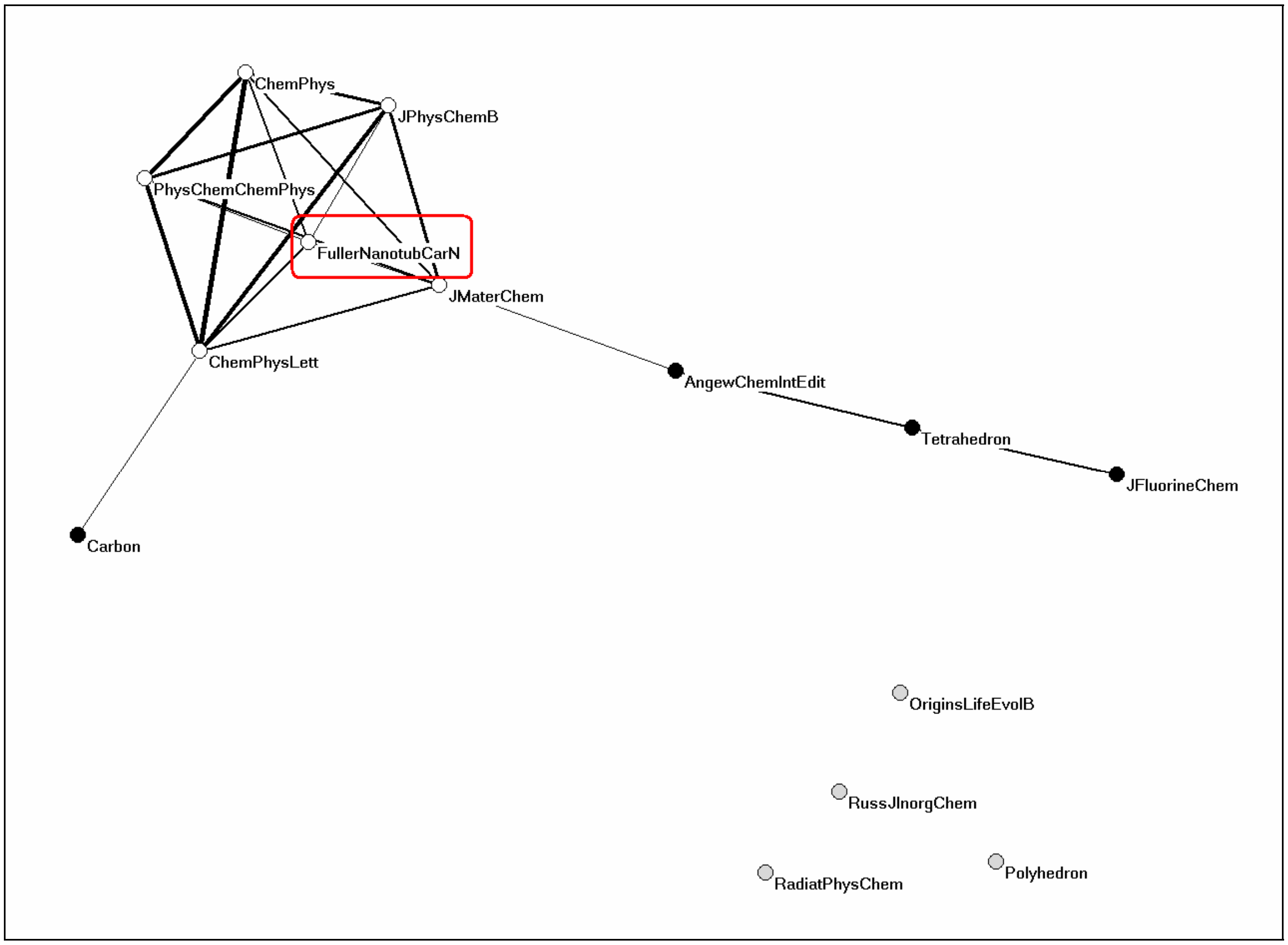

**Figure 1**: Citation Impact Environment of 14 journals citing *Fullerenes Nanotubes and Carbon Nanostructures* more than once (cosine ≥ 0.2).

Figure 1 shows that *Fullerenes Nanotubes and Carbon Nanostructures* is firmly integrated in a set of chemical-physics journals. However, we shall see below that the journals citing and cited by this journal are heavily interwoven with the journals in the environment of the other five journals. Let us first combine the sets of journals citing or cited by these six journals.

*3.1 The citation environment of the combined set*

The local citation environments of the core journals in nanotechnology are sometimes very large. For example, *Nano Letters* is cited by articles in 305 journals more than once.[5] However, only 17 of these journals cite *Nano Letters* to the extent of more than one percent of its total citation rate of 7,349. Authors in *Nano Letters* themselves cite 372 journals, of which only 16 to the extent of more than one percent of the journal's total references (12,131). In order to discard these large tails of the distributions, we shall use this one-percent threshold for the delineation (He & Pao, 1986; Leydesdorff & Cozzens, 1993).

Thirty-seven journals *cite* one of the six seed journals with the stem "nano" in their title above the threshold, and 53 journals are *cited* by them. Since there is an overlap of 23 journals among these two subsets, 67 journals can be considered as "nano-relevant" journals. Using the same threshold of one percent, Zhou & Leydesdorff (2006) found 85 journals to be "nano-relevant" in 2003 using only three instead of six seed journals. Therefore, the conclusion seems justified that the nano-relevant environment among scientific journals is an increasingly focused set.

[5] All single citations are aggregated by the ISI and subsumed under the category "All others".

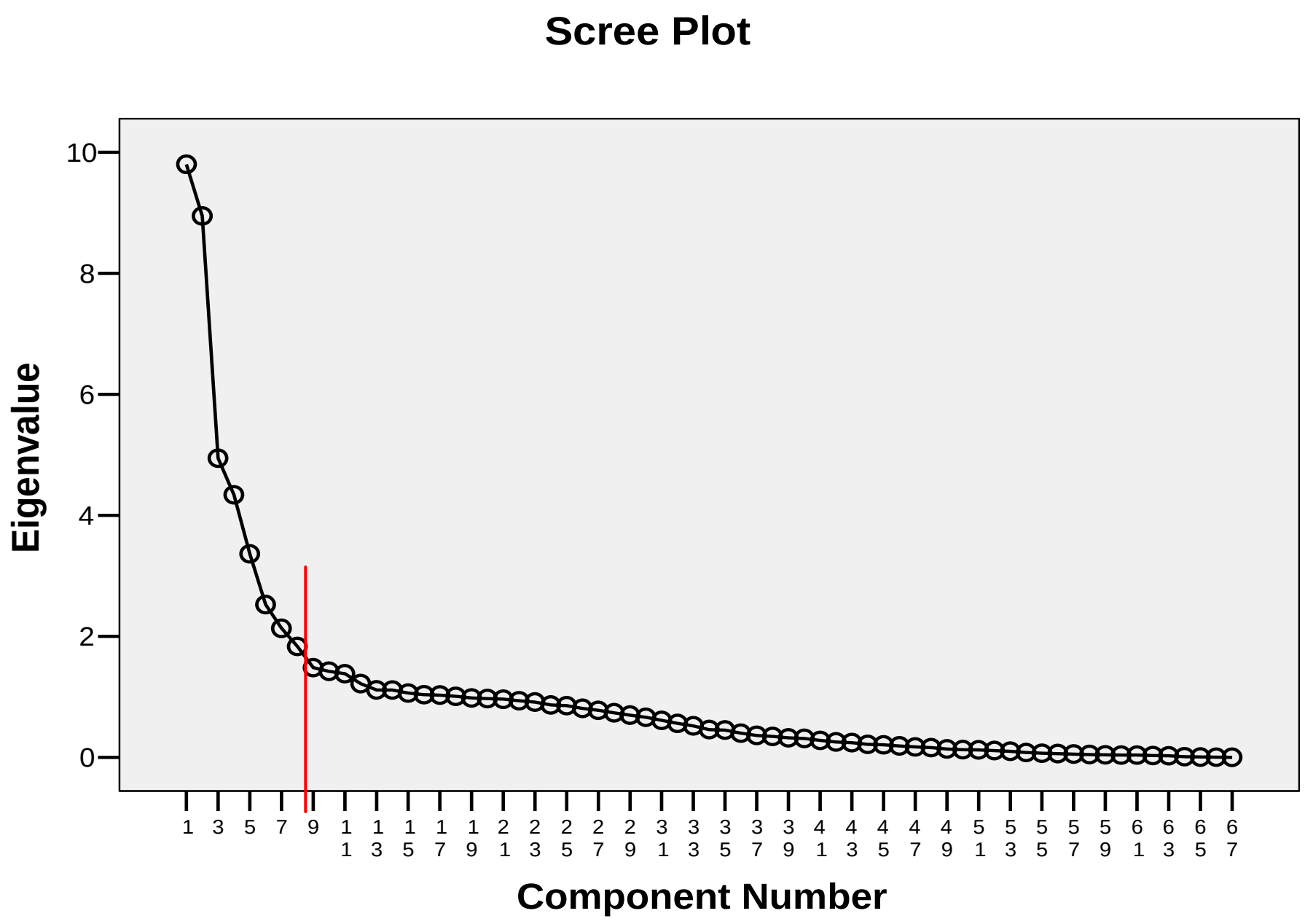

**Figure 2**: Screeplot of 67 journals in the relevant citation environment of six core nano-journals.

When the citation matrix among these 67 journals is analyzed in terms of being-cited patterns, the screeplot (Figure 2) suggests the extraction of eight factors explaining 56.5% of the variance in this matrix. Table 3 provides the eight-factor solution using Varimax rotation and Kaiser normalization.

**Rotated Component Matrix(a)**

| | Component | | | | | | | |
|---|---|---|---|---|---|---|---|---|
| | 1 | 2 | 3 | 4 | 5 | 6 | 7 | 8 |
| SOLID STATE ELECTRON | .868 | -.115 | -.122 | | | | | |
| IEEE ELECTR DEVICE L | .811 | -.121 | -.169 | -.182 | | | -.185 | |
| P IEEE | .788 | -.168 | -.109 | | | | -.164 | |
| APPL PHYS LETT | .777 | | .113 | .392 | | | .169 | |
| IEEE T ELECTRON DEV | .751 | -.119 | -.186 | -.186 | | | -.217 | .107 |
| J APPL PHYS | .748 | | | .452 | | | .233 | |
| JPN J APPL PHYS | .658 | | | | | | .148 | |
| J VAC SCI TECHNOL B | .632 | | | .120 | | | | |
| SEMICOND SCI TECH | .629 | | | .562 | | | | |
| MRS BULL | .612 | | .459 | .387 | | .112 | .380 | |
| **IEEE T NANOTECHNOL** | **.557** | **-.128** | **.149** | **.193** | | | **-.183** | **-.111** |
| APPL SURF SCI | .417 | -.102 | .181 | .294 | | -.146 | .176 | -.158 |
| J CRYST GROWTH | .334 | | | .103 | | | .248 | |
| **FULLER NANOTUB CAR N** | **-.180** | | | | | **-.113** | | **-.166** |
| AEROSOL SCI TECH | -.124 | | | | | | | |
| POWDER TECHNOL | -.113 | | | | | | | |
| RUSS J INORG CHEM+ | -.108 | | | | | | | |
| J ORG CHEM | | .925 | -.124 | | | | | |
| TETRAHEDRON | | .914 | -.149 | | | | | |
| CHEM COMMUN | | .911 | .259 | | | .100 | | |
| TETRAHEDRON LETT | | .906 | -.151 | | | | | |
| CHEM LETT | | .853 | .205 | | | | | |
| CHEM REV | | .848 | .248 | | .337 | .162 | | .101 |
| J AM CHEM SOC | | .775 | .296 | | .290 | .230 | | |
| ANGEW CHEM INT EDIT | | .758 | .172 | | | .198 | | |
| J FLUORINE CHEM | -.101 | .403 | -.102 | | | -.127 | | |
| POLYHEDRON | -.100 | .232 | | | | | | |
| RUSS CHEM B+ | -.108 | .119 | | | | | | |
| *ADV MATER* | *.111* | *.215* | *.840* | | | *.102* | | *.230* |
| ***NANO LETT*** | ***.130*** | | ***.796*** | | | ***.159*** | | |
| *CHEM MATER* | | *.184* | *.788* | | | | *.148* | *.234* |
| ***J NANOPART RES*** | | | ***.737*** | | | | ***.132*** | ***-.163*** |
| *LANGMUIR* | | | *.725* | | *.108* | | *-.105* | *.163* |
| *J MATER CHEM* | | *.247* | *.715* | | | | *.172* | *.131* |
| ***J NANOSCI NANOTECHNO*** | ***.138*** | | ***.703*** | ***.183*** | | | | |
| *J PHYS CHEM B* | | *.110* | *.676* | | *.504* | | | |
| *J COLLOID INTERF SCI* | *-.106* | | *.554* | | | | | *.103* |
| ***NANOTECHNOLOGY*** | ***.359*** | | ***.454*** | ***.356*** | | | | ***-.106*** |
| *CARBON* | | | *.207* | | | | | *-.128* |
| *ANAL CHEM* | | | *.157* | | | | *-.100* | |
| PHYS REV B | .135 | | | .923 | | | | |
| J PHYS-CONDENS MAT | .118 | | | .898 | .126 | | | |
| PHYS REV LETT | | | | .881 | | .140 | | |
| PHYSICA E | .306 | | | .785 | | | | |
| PHYS SOLID STATE+ | | | | .605 | | | | |
| SURF SCI | | | .238 | .541 | .258 | | | |
| PHILOS T ROY SOC A | | -.120 | | .515 | .293 | .194 | | |
| APPL PHYS A-MATER | .447 | | .273 | .468 | | | .152 | -.187 |
| CHEM PHYS | | | | .109 | .961 | | | |
| CHEM PHYS LETT | | | .128 | .115 | .954 | | | |
| J CHEM PHYS | | | | .121 | .911 | | | |
| J PHYS CHEM A | | .129 | | | .897 | | | |
| PHYS CHEM CHEM PHYS | | | .256 | | .855 | | | |
| P NATL ACAD SCI USA | | | | | | .884 | | |
| SCIENCE | | .106 | .247 | .322 | .146 | .859 | | |
| NATURE | | | .120 | .400 | | .852 | | |
| ORIGINS LIFE EVOL B | | | | | | .327 | | |
| J MATER SCI | | | | | | | .793 | .183 |
| J AM CERAM SOC | | | | | | | .758 | |
| J MATER RES | .315 | | .146 | .103 | | | .683 | |
| J ENG MATER-T ASME | | -.103 | -.124 | | | | .398 | |
| IEEE T VLSI SYST | | | | | | | -.193 | |
| WEAR | | | | | | | .187 | |
| J KOREAN PHYS SOC | | | | | | | | |
| POLYMER | | | .118 | | | | | .864 |
| MACROMOLECULES | | | .225 | | | | | .862 |
| RADIAT PHYS CHEM | | | | | .114 | | | .153 |

Extraction Method: Principal Component Analysis.
Rotation Method: Varimax with Kaiser Normalization.
a Rotation converged in 7 iterations.

**Table 3**: Factor solution of the being-cited patterns of 67 nano-relevant journals citing or cited by six core journals in nanotechnology (factor loadings ≥ 0.1).

The factor solution is very clear and can be designated in terms of the relevant disciplines. Four of the six seed journals have their primary factor loading on Factor 3 (which explains 7.4% of the variance in the matrix). However, *IEEE Transactions on Nanotechnology* belongs to a first group of journals with loadings on Factor 1 (14.6% of the variance). *Fullerenes Nanotubes and Carbon Nanostructures* does not load positively on any of the factors extracted and can thus be considered as an isolate.

All six seed journals exhibit considerable factorial complexity, but within this factor matrix factorial complexity is not an exclusive property of these journals (Van den Besselaar & Heimeriks, 2001). For example, the *IEEE Electron Device Letters* has the highest interfactorial complexity in this matrix (with factor loadings on six of the eight factors). However, this journal has a negative loading on the factor which was designated as “nano” (Factor 3); it belongs to a different group.

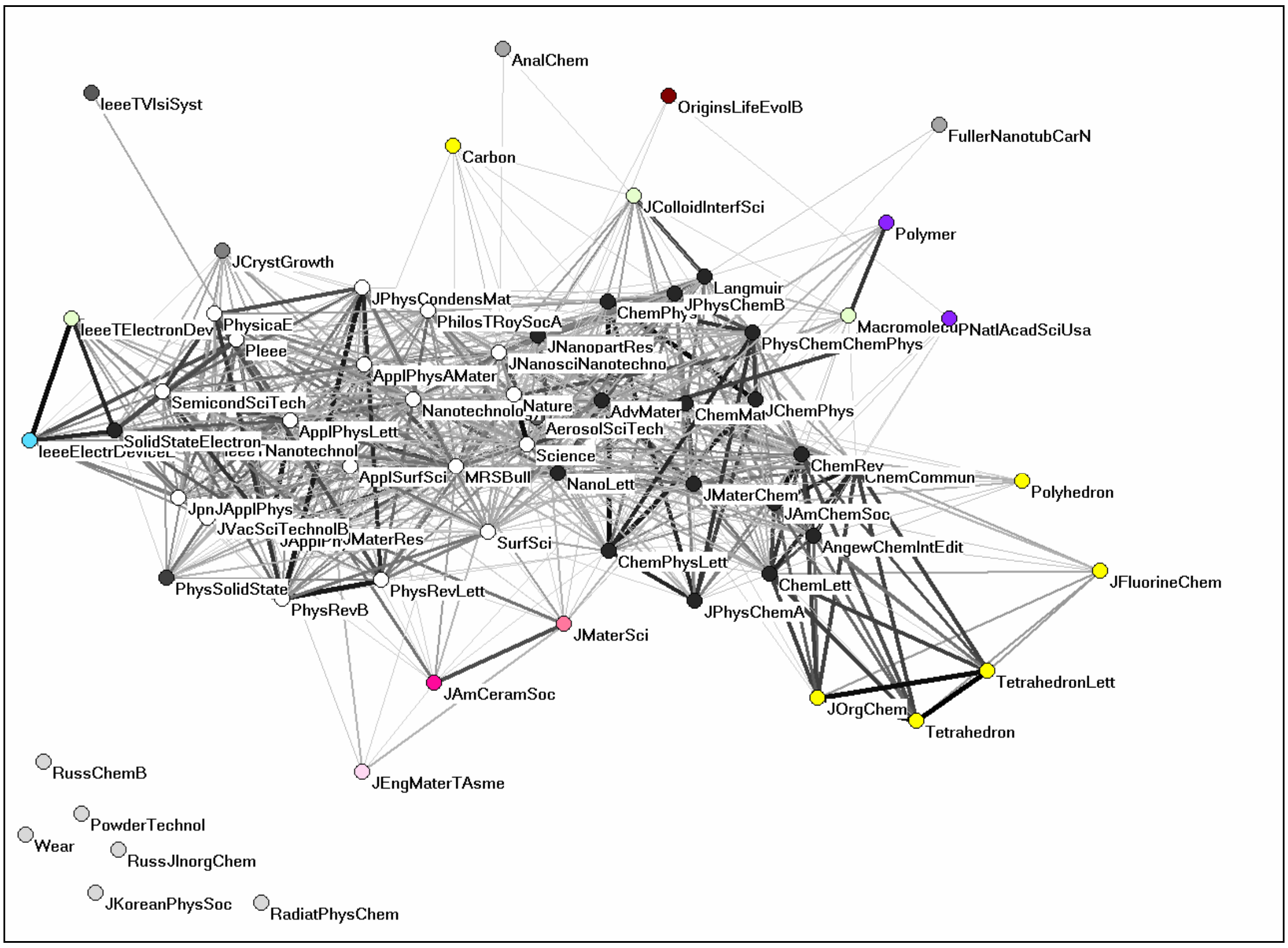

**Figure 4**: Visualization of the local citation impact of 67 nano-relevant journals citing six core journals in nanoscience and nanotechnology (cosine ≥ 0.2).

Figure 4 shows that without the rotation of the factor analysis, the structure among the nano journals is overshadowed by their disciplinary affiliations. The core algorithm available in Pajek organizes the chemistry journals into one cluster (with dark vertices) and a physics cluster into another (white vertices). *Nano Letters* and *Nanoparticle Research* are attributed to the chemistry cluster, while *Nanotechnology, IEEE Transactions on Nanotechnology*, and the *Journal of Nanoscience and Nanotechnology* are part of the physics cluster. *Fullerenes Nanotubes and Carbon*

*Nanostructures* is positioned at the top right of the figure as a special case with connections to specific journals in physical chemistry.

In summary, the nano core-set is positioned at an interface between chemistry and physics, and highly interwoven with general science journals. Among the latter, *Science* and *Nature* are present at the same interface as the nano-journals. Note that *Science* and *Nature* are both attributed to the physics set in this context, while the *PNAS* is positioned on the side of chemistry.

Disciplinary journals like the *Journal of the American Chemical Society, Physics Review Letters,* and *Physics Review B* are also part of this environment. These journals do not publish only or even mainly nano-relevant literature. Thus, while we have been able to zoom in on a specific set, we have not yet been able to delineate a strictly nano-relevant set (Leydesdorff, 2006b).

### *3.2. "Betweenness centrality" as a measure for interdisciplinarity*

In another context, one of us has proposed using "betweenness centrality" as a measure of interdisciplinarity (Leydesdorff, 2006a). "Betweenness centrality" is a measure of how often a node (vertex) is located on the shortest path (geodesic) between other nodes in the network (Freeman, 1977; 1978/1979).[6] If a node with a high level of betweenness were deleted from a network, the network would fall apart

[6] If $g_{ij}$ is defined as the number of geodesic paths between $i$ and $j$ and $g_{ikj}$ is the number of these geodesics that pass through $k$, $k$'s betweenness centrality is defined as (Farrall, 2005):

$$\sum_i \sum_j \frac{g_{ikj}}{g_{ij}}, \quad i \neq j \neq k$$

into otherwise coherent clusters. Betweenness is normalized by definition as the proportion of all geodesics that include the vertex under study and can thus be expressed as a percentage.

If one applies this centrality measure (available in Pajek) to the representation provided in Figure 4, one obtains the representation in Figure 5. The figure shows the size of the vertices proportional to their respective “betweenness centrality.” It highlights the nano-core journals more specifically than the rest, but the values for journals with a scope broader than nanotechnology sometimes remain similar to those of nano-core journals.

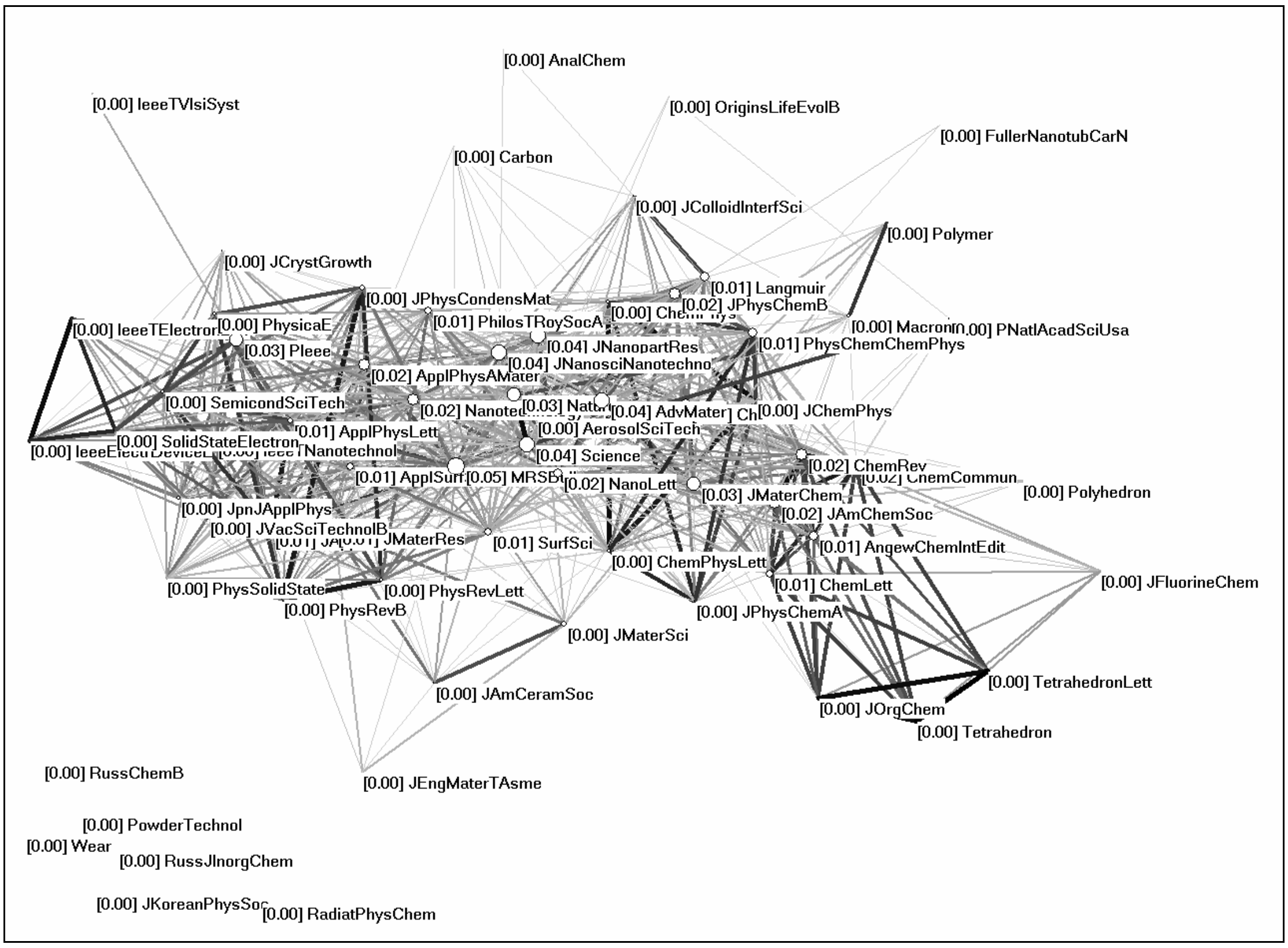

**Figure 5**: Betweenness centrality among the 67 nano-relevant journals (cosine ≥ 0.2).

For example, the *MRS Bulletin* has a betweenness centrality of 5.27%, and the *Journal of Nanoscience and Nanotechnology* of 4.09%, but *Science* scores with 4.07% before other leading nanotechnology journals. For example, *Nanotechnology* has only 2.24% betweenness. Thus, the group with high betweenness centrality includes journals that are not specific for nanotechnology.

### *3.3. Cited versus citing*

The difference between the being-cited environment of the core set and the citing environment may be another factor relevant for the delineation. In a previous study, Leydesdorff *et al.* (1994) argued that new developments can be traced in the being-cited environments first because new developments (e.g., discoveries) can be expected to draw the attention of authors in neighboring areas. Authors in these areas may begin to cite from the new journals. In this study, however, we are not interested in "nanotechnology" as a completely new development because it was already established as a field during the 1990s (Braun *et al.*, 1997; Meyer & Persson, 1998). For example, a Nobel Prize was given in 1996 to Robert Curle, Harald Kroto, and Richard Smalley for the discovery of "buckyballs," one of the fullerenes (Kroto *et al.*, 1985).

Thus, our research question here is not whether a new development has been noted by researchers in surrounding fields, but whether this new techno-science has developed to such an extent that a specific set of journals can be delineated from journals in the relevant environments. One would expect authors who publish in the new journals to be the first to draw a distinction in their publications and citations between a core set and journals in relevant environments. Do researchers active in the emerging field feel inclined to change their publication and citation practices? (Gilbert, 1977; Small, 1978; Leydesdorff & Van der Schaar, 1987). This change in practice would be visible on the active "citing"-side of the database more than on the "cited"-side.

Indeed, the representation of the vector for betweenness centrality among the 53 journals citing the six seed journals does not improve on the representation in Figure 5, but the representation of the vector-space among the 38 journals cited by the six seed journals is considerably different (Figure 6).[7]

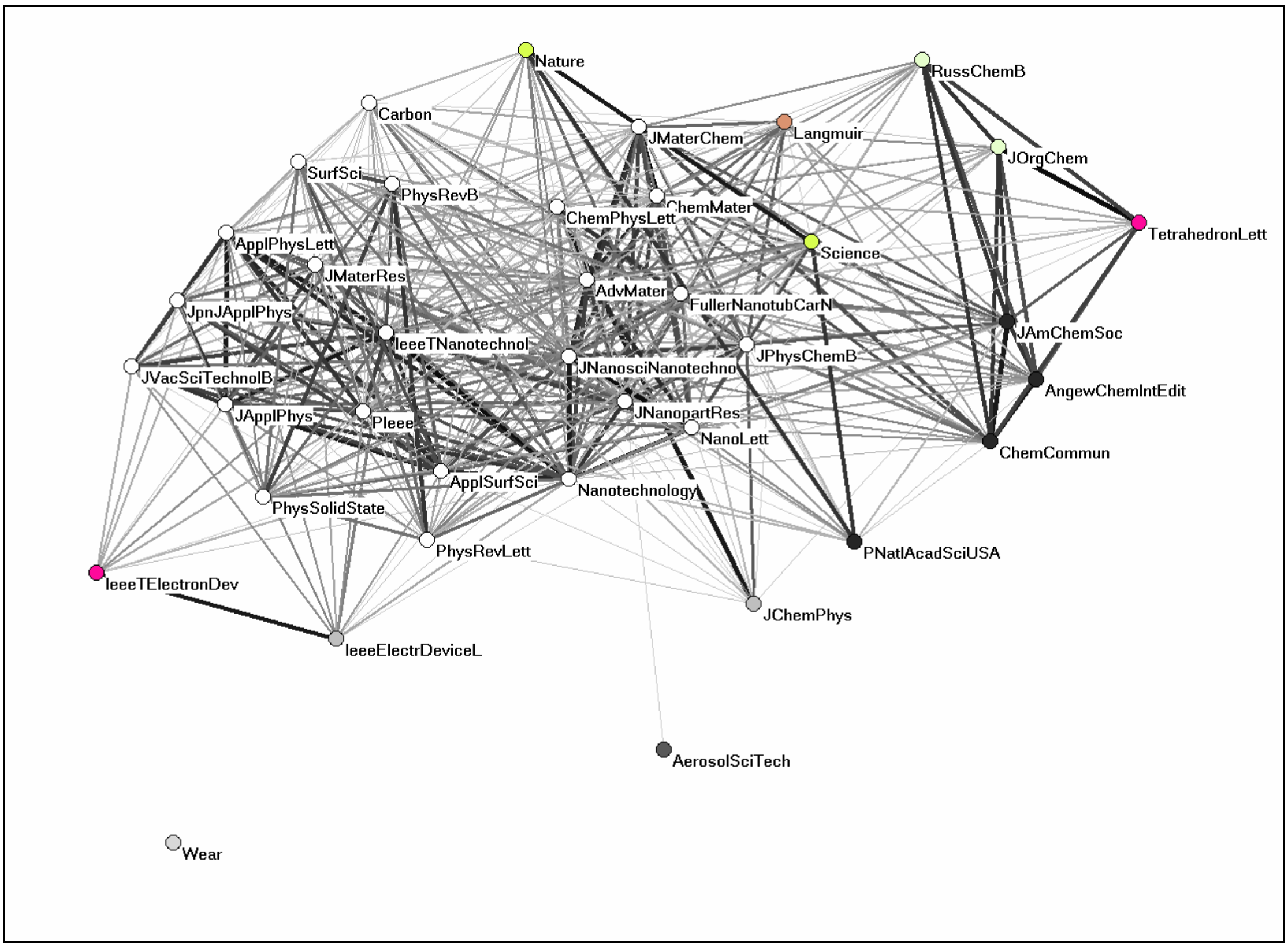

**Figure 6**: Partitioning of citing patterns among 38 journals cited by the six seed journals in nanotechnology (cosine ≥ 0.2).

[7] The *Journal of Nanoscience and Nanotechnology* has to be added to this list of originally 37 journals because this journal is not included when the threshold is set at one percent. For example, the number of within-journal "self"-citations by this journal in 2004 is only 30 while the 1% threshold is 56.

The six nanotechnology journals are positioned in the middle of the figure, with certain other journals (like *Chemistry of Materials*) as a relatively separate cluster. *Science, Nature,* and *PNAS* are nearby, but differently attributed. However, neither the algorithm available in Pajek nor factor analysis is able to distinguish the grouping of these nano-journals from the physics set (which is now more dominant than the chemistry one).

Betweenness centrality shows a distinctively different pattern for the ten journals within the ellipse (Figure 7). Note that these journals include *Fullerenes Nanotubes and Carbon Nanostructures* but not the *IEEE Transactions on Nanotechnology.* The latter journal is found more deeply in the physics cluster.

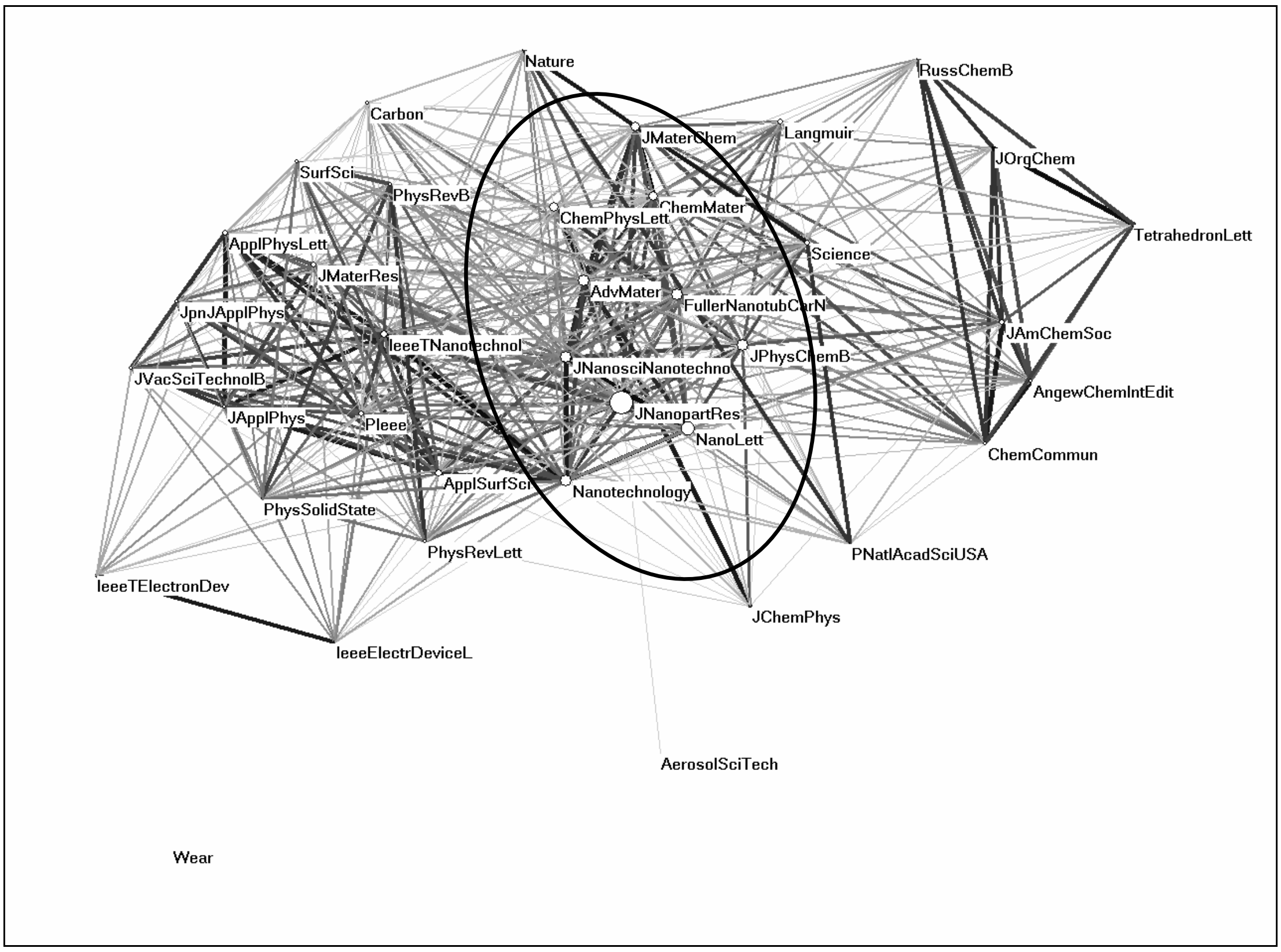

**Figure 7**: Betweenness centrality among 38 journals cited by the six nano-journals to the extent of more than one percent of their citation totals (cosine ≥ 0.2).

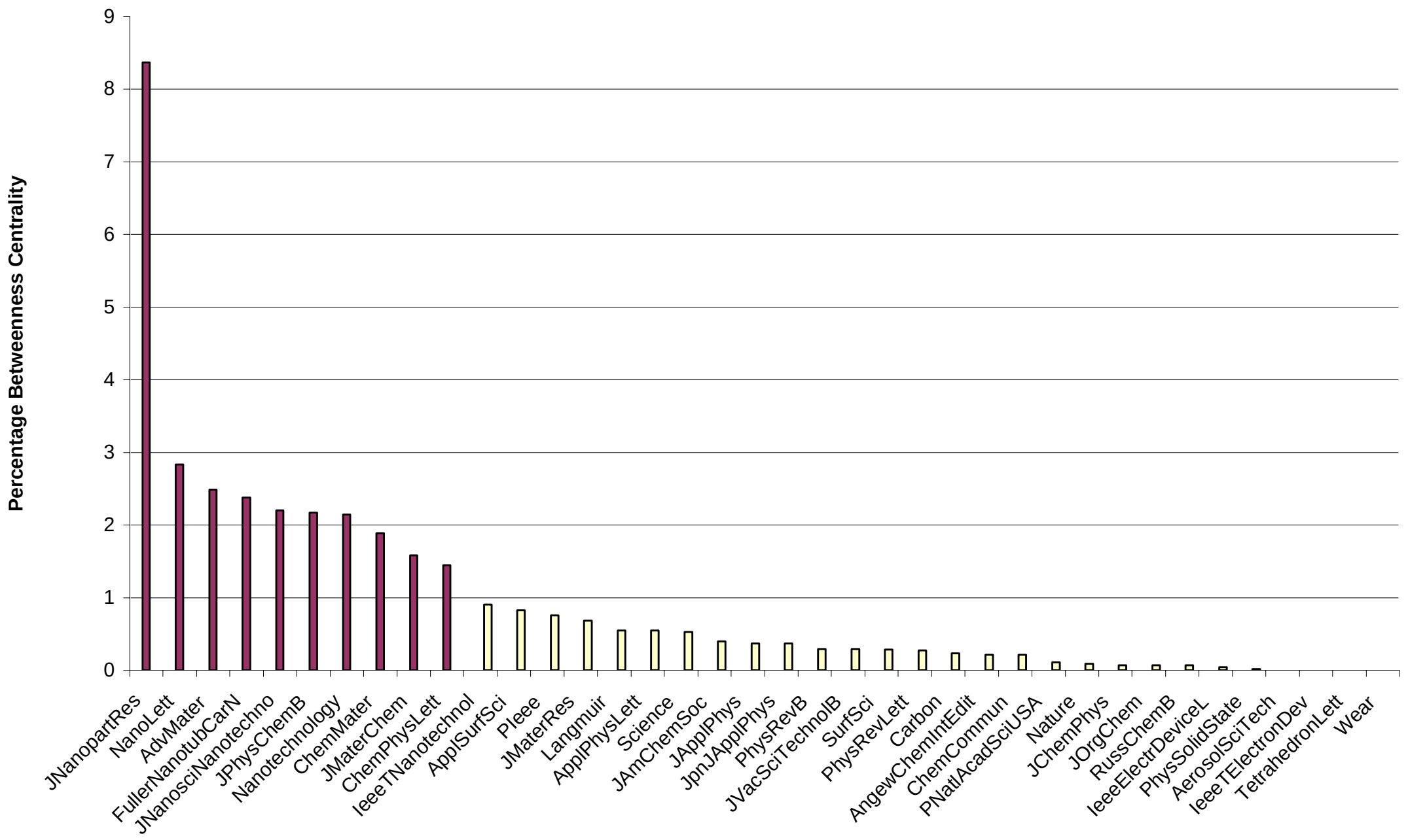

**Figure 8**: Percentage betweenness centrality for 38 journals constructing nanotechnology in terms of aggregated citation patterns (citing).

Figure 8 provides the betweenness centrality values which correspond to the delineations penciled into Figure 7. Table 4 provides further evidence for the difficulty in identifying this emerging cluster otherwise by using factor analysis of this same matrix.[8] The ten journals are boldfaced in this matrix. Their factor loadings are concentrated in Factor 2 and exhibit factorial complexity, but not all to the same degree. *Nanotechnology*, for example, is placed outside the relevant grouping, while *Langmuir* is placed within it.

**Rotated Component Matrix(a)**

| | Component | | | | | |
|---|---|---|---|---|---|---|
| | 1 | 2 | 3 | 4 | 5 | 6 |
| APPL PHYS LETT | .866 | .203 | | .324 | | |
| J APPL PHYS | .829 | .139 | | .419 | | |
| J VAC SCI TECHNOL B | .822 | .106 | | | | |
| IEEE T NANOTECHNOL | .805 | .123 | | .486 | .240 | |
| JPN J APPL PHYS | .797 | | | .152 | | |
| P IEEE | .729 | | | | .343 | |
| IEEE ELECTR DEVICE L | .717 | -.266 | | -.299 | | .130 |

[8] A six factor solution explaining 72.5% of the variance is suggested by inspection of the screeplot.

| | | | | | | |
|---|---|---|---|---|---|---|
| APPL SURF SCI | .695 | .203 | | .535 | -.110 | |
| **NANOTECHNOLOGY** | **.623** | **.421** | | **.526** | **.320** | |
| IEEE T ELECTRON DEV | .613 | -.289 | | -.302 | | .149 |
| J MATER RES | .602 | .304 | -.106 | .165 | | |
| ***CHEM MATER*** | | ***.827*** | ***.347*** | | | |
| ***J MATER CHEM*** | | ***.815*** | ***.358*** | | ***.163*** | |
| ***ADV MATER*** | ***.295*** | ***.774*** | ***.171*** | | ***.345*** | |
| ***J NANOPART RES*** | ***.306*** | ***.762*** | | ***.279*** | | ***.132*** |
| ***J NANOSCI NANOTECHNO*** | ***.361*** | ***.712*** | | ***.345*** | ***.348*** | ***.153*** |
| *LANGMUIR* | *-.112* | *.687* | | | | *.192* |
| ***NANO LETT*** | ***.329*** | ***.570*** | ***.144*** | ***.308*** | ***.533*** | ***.111*** |
| J ORG CHEM | -.109 | | .923 | | | |
| CHEM COMMUN | -.128 | .360 | .877 | | | |
| TETRAHEDRON LETT | -.116 | | .864 | | | |
| RUSS CHEM B+ | -.152 | | .860 | | | |
| J AM CHEM SOC | | .403 | .804 | | | .218 |
| ANGEW CHEM INT EDIT | -.107 | .255 | .801 | | | |
| AEROSOL SCI TECH | | | -.145 | | | |
| WEAR | -.100 | | -.133 | | -.103 | -.127 |
| PHYS REV B | .233 | | | .890 | | |
| PHYS REV LETT | .143 | | | .755 | .215 | .137 |
| PHYS SOLID STATE+ | .251 | | | .750 | | |
| SURF SCI | | | | .712 | -.116 | .211 |
| **FULLER NANOTUB CAR N** | | **.114** | **.355** | **.594** | **.249** | **.219** |
| CARBON | | .142 | | .288 | .228 | |
| NATURE | | .112 | | .119 | .919 | |
| SCIENCE | | .190 | | | .917 | |
| P NATL ACAD SCI USA | | | | | .844 | |
| J CHEM PHYS | | | | .139 | | .916 |
| **CHEM PHYS LETT** | | **.177** | **.100** | **.238** | | **.896** |
| **J PHYS CHEM B** | | **.567** | **.204** | **.181** | | **.689** |

Extraction Method: Principal Component Analysis.
Rotation Method: Varimax with Kaiser Normalization.
a Rotation converged in 8 iterations.

**Table 4**: Factor solution of the citing patterns of 38 nano-relevant journals cited by six seed journals in nanoscience and nanotechnology (factor loadings ≥ 0.1).

Table 5 provides the Library of Congress information for the ten core journals thus discerned (Bensman, *personal communication*). The table illustrates the difficulty with hierarchical indexes (Bensman, forthcoming). Some of the chemistry journals are not classified as nano-journals using this index. In other words, none of the other available methods (multivariate analysis; inductive classification) enabled us to distinguish the core group of ten nanotechnology journals emerging in the database.

| Journal Title | LC Subject Headings | LC Class Number | LC Class Name | LC Class Hierarchy |
|---|---|---|---|---|
| Chemistry of materials | 1) Chemistry. 2) Materials. | QD1 | Chemistry | Chemistry |
| The journal of physical chemistry. B, Condensed matter, materials, surfaces, interfaces & biophysical | 1) Chemistry, Physical and theoretical. | QD1 | Chemistry | Chemistry |
| Fullerenes, nanotubes, and carbon nanostructures. | 1) Fullerenes. 2) Nanotubes. 3) Nanostructures. | QD181.C1 | Special elements: Carbon | Chemistry—Inorganic chemistry--Special elements. By chemical symbol, A-Z |
| Chemical physics letters | 1) Chemistry, Physical and theoretical. | QD450 | Physical and theoretical chemistry | Chemistry—Physical and theoretical chemistry |
| Journal of nanoscience and nanotechnology | 1) Nanoscience. 2) Nanotechnology. | T174.7 | Nanotechnology | Technology (General)—Nanotechnology |
| Nano letters. | 1) Nanotechnology. | T174.7 | Nanotechnology | Technology (General)—Nanotechnology |
| Nanotechnology. | 1) Nanotechnology. | T174.7 | Nanotechnology | Technology (General)—Nanotechnology |
| Advanced materials | 1) Materials. 2) Chemical vapor deposition. | TA401 | Materials of engineering and construction | Engineering (General). Civil engineering (General)--Materials of engineering and construction |
| Journal of materials chemistry. | 1) Materials science. 2) Materials | TA401 | Materials of engineering and construction | Engineering (General). Civil engineering (General)--Materials of engineering and construction |
| Journal of nanoparticle research | 1) Nanoparticles. | TA418.78 | Materials as particles, with tests | Engineering (General). Civil engineering (General)--Materials of engineering and construction--Physical properties--Materials as particles, with tests |

**Table 5**: Ten core journals of nanoscience and nanotechnology in the classification of the U.S. Library of Congress.

On the assumption that the ten journals listed form a core set for nanoscience and technology journals, one can make a selection from the *Science Citation Index* and count, for example, country addresses. The result at the level of nations is provided in Figure 9.

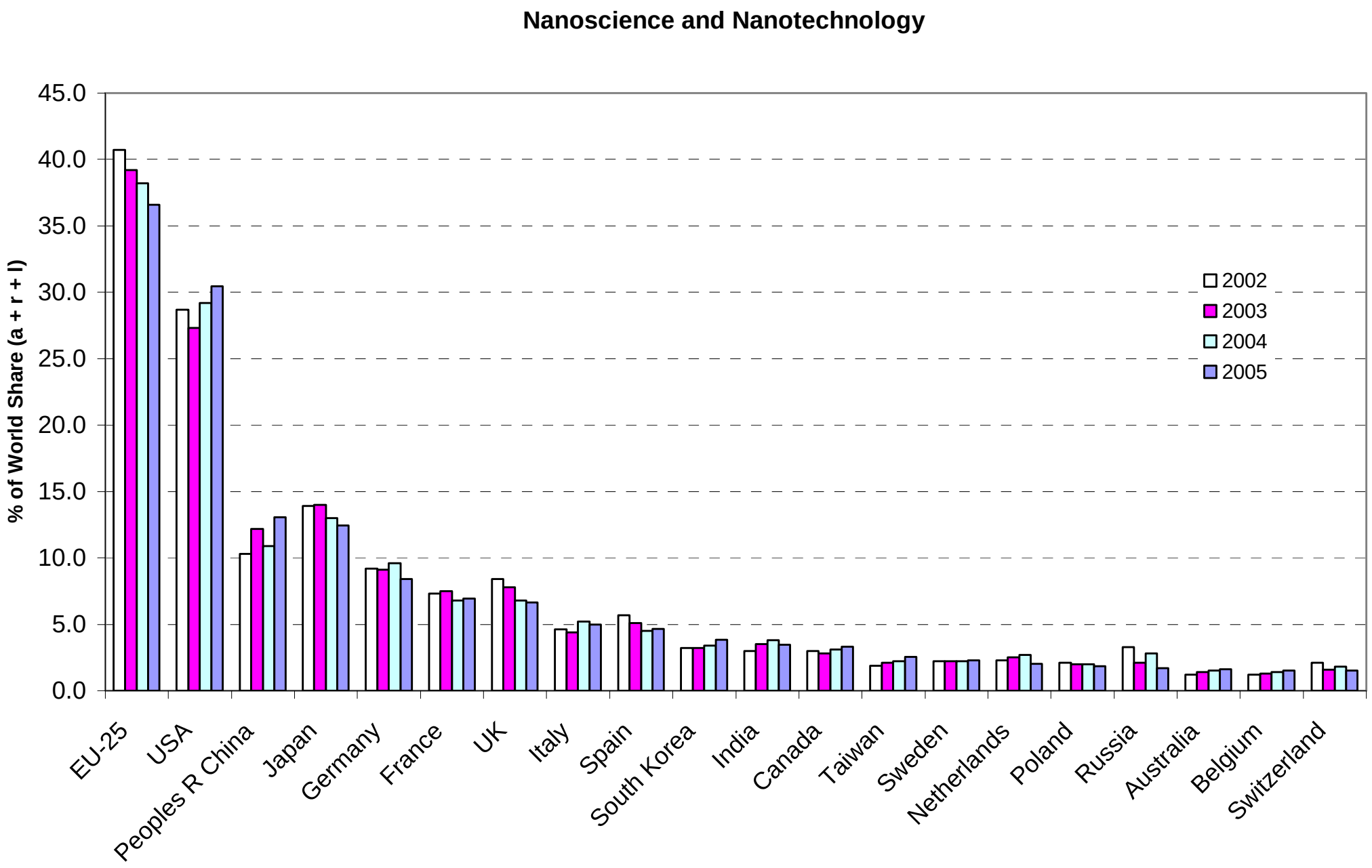

**Figure 9**: Percentage of world share of publications in ten core journals in nanotechnology (2003-2005) for seventeen leading countries (integer counting).

Within this set, the EU-25 is loosing each year more than one percent of its world share of publications. As a nation, the position of the U.S.A. is unambiguously the first; the percentage of contributions with an American address is increasing. Recently, China obtained the second position, while Japan is loosing "market" share (Kostoff, 2004; Zhou & Leydesdorff, 2006). The order among the major players is rather stable, but there are shifts in the order of half a percentage point during these four years of observation. For example, Taiwan has improved its position from 1.9% in 2002 to 2.6% in 2005.

## 4. US Patent data

Since nanotechnology is defined more as a technology than as a science, we wondered whether patent data might provide us with a better indicator of the relevant journal set by using the non-patent literature references (NPLR) within the patents. The NPLRs may contain the names of scientific journals. For this purpose, we downloaded the 336 patents classified by the USPTO as nanotechnology (Class 977) during 2005. The search string was: “CCL/977/$ and ISD/$/$/2005”.

These patents contain 1,948 NPLR, of which we could use 1,146 with a hundred names of scientific journals. Figure 10 provides the distribution in a pie-chart format.

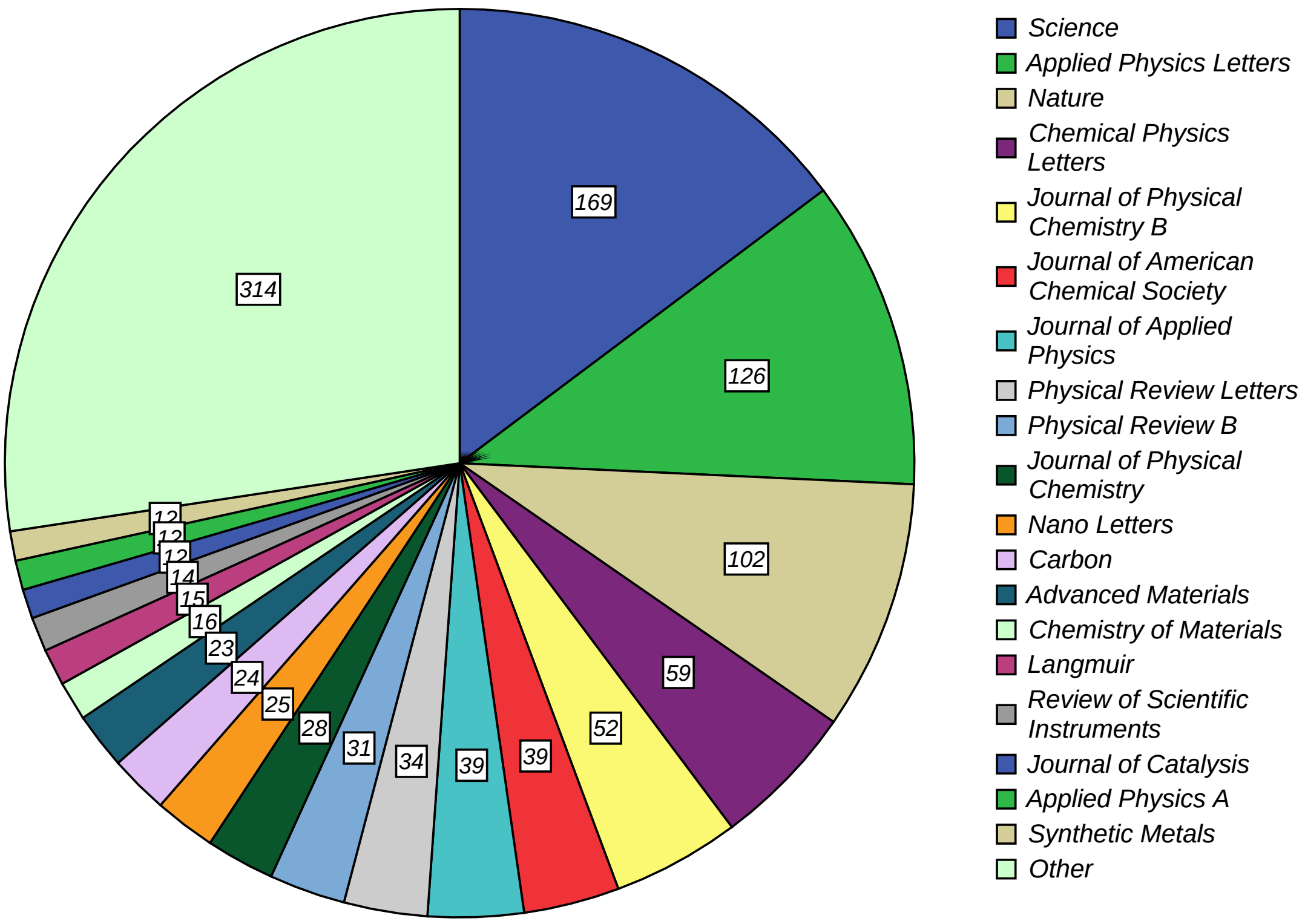

**Figure 10**: Distribution of journals cited in the NPLR of patents within the nanotechnology class during 2005.

The conclusion is that the references to the scientific knowledge base of the patents (Leydesdorff, 2004) are not specific enough for the delineation of a core set of nano-journals. The first four journals are *Science, Nature,* and two journals that publish letters. Among the latter two is *Chemical Physics Letters*, which was included above among the ten core journals. However, *Applied Physics Letters*—the other journal--was not classified above as a core journal.

The *Journal of Physical Chemistry B* follows on the fifth position, but *Nano Letters*—as the first journal with "nano" in its title—follows only at the eleventh position with 25 references. For the purpose of delineating a journal set within the domain of the *Science Citation Index*, patents thus do not seem of much help (Hedge & Sampat, 2005; Sampat, 2006, at p. 784, note 28; Meyer, forthcoming). We don't expect that European or Japanese patents would perform much better in this respect except that they would, of course, control for the regional bias in the U.S. database (Narin *et al*. 1997).

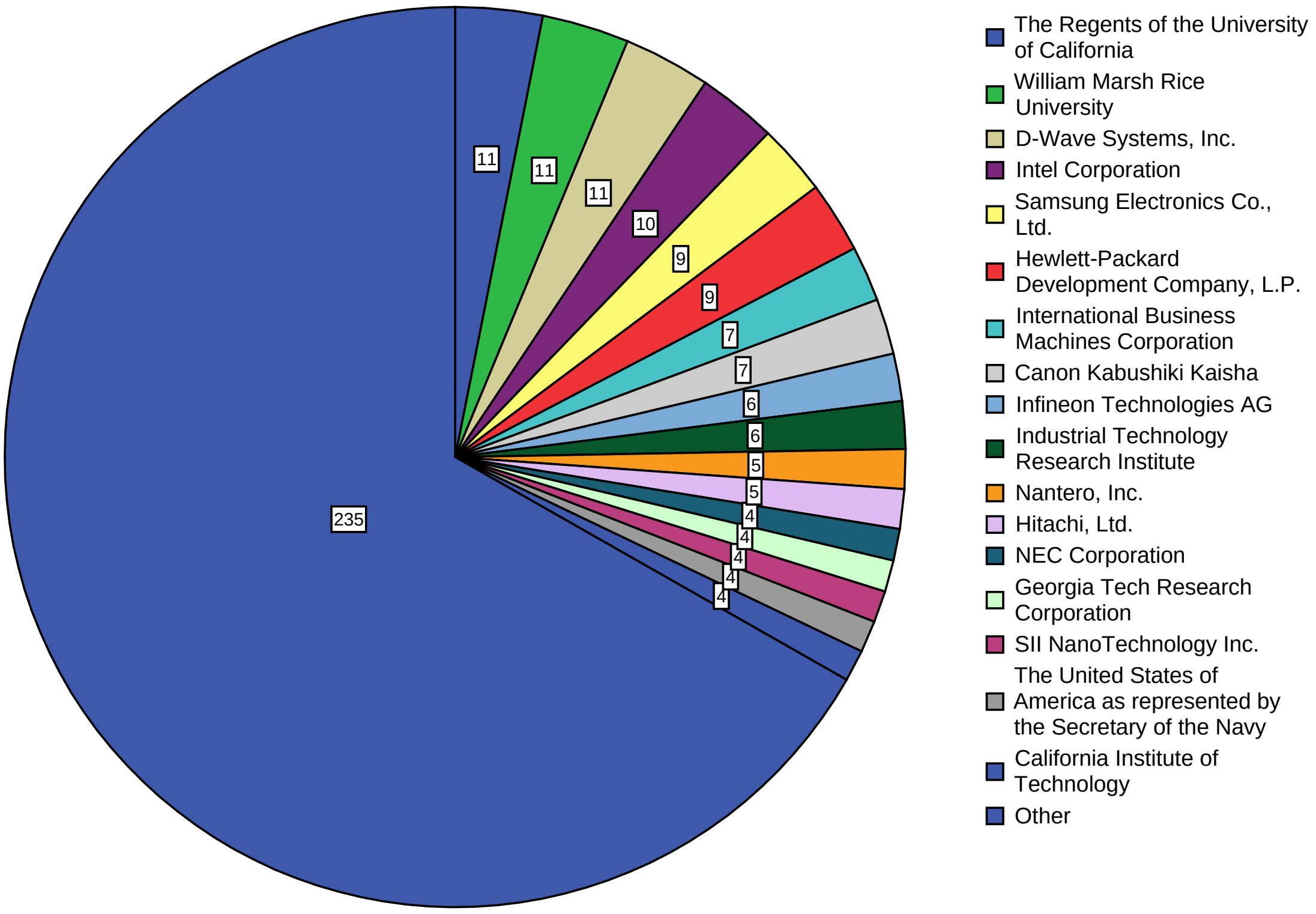

**Figure 11**: Assignees in descending order with more than one percent of the 336 nano-patents in 2005.

The geographical distribution of the U.S. patents can be evaluated both in terms of assignees and inventors. There are 352 assignees to these 336 patents, and 1027 inventors. This confirms that intellectual property is mostly unambiguous, but that invention is increasingly based on collaborations. It goes beyond the scope of this study to analyze these co-inventions in terms of triple-helix relations.

Among the assignees (Figure 11), the central position of the Regents of the University of California is not unexpected, because the various branches of the University of California all patent under this heading. The William Marsh Rice University (Houston, Texas) impresses with an equal number of eleven patents on its own. Fullerenes were discovered at this university, leading to the Nobel Prize in 1996 (Kroto *et al.*, 1985). The other major holders of patents are all corporations and technological institutes.

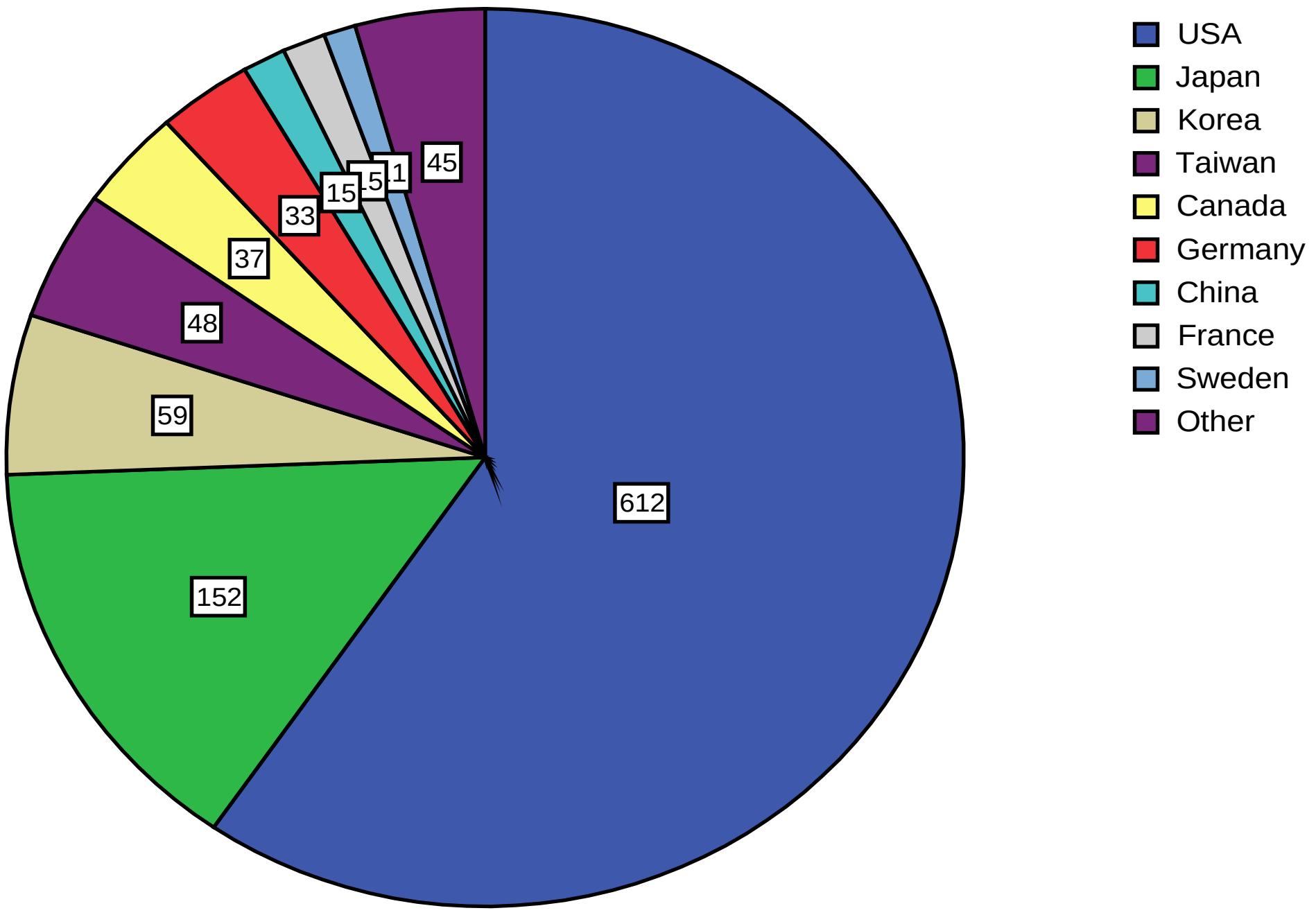

**Figure 12**: Regional origin of the 1027 inventors of the 336 nano-patents.

Figure 12 shows the regional distribution of the 1027 inventors of these patents. Although the leading Asian nations are represented, the American origin of the database is evident. Among the European countries, only 33 German, 15 French, and 11 Swedish addresses of inventors are notable (given the 1% threshold of the category "Other").

It is tempting to pursue the analysis further in terms of differences in repertoires (e.g., co-word patterns) between the patent set and the journal-article set, but this would lead us astray from the purpose of this study and the original research question. Let us therefore summarize and draw conclusions.

## 5. Conclusions

Our research question was whether it was possible using aggregated citation data among journals and/or patents to delineate a specific nanotechnology set of scientific journals. Using the USPTO data, it became clear that the references in the class of patents specifically designated as nanotechnology were too general for this delineation. One can expect that similar patterns would emerge in European and Japanese sets of patents. Furthermore, patents are biased in terms of the world region which they cover. For example, we found few European addresses among the more than 40% non-American inventors in the nano-class.[9]

This leaves us with the journal literature. The emergence of a new development is first noted on the cited side (Leydesdorff *et al.*, 1994), but our research question was whether the codification in an emerging field of science would be strong enough to make delineation possible. One would expect this codification of the repertoire to be strongest among practicing scientists who, under the influence of the emerging specialty, gradually change their publication and citation behaviour. We have reason to believe that this process is ongoing:

- Based on 2004 data, we found that the relevant citation environment using six core nano-journals consisted of 67 journals, while a year earlier we found 85 journals in the environment using only three core journals with a similar threshold of one percent. Thus, the relevant environment is shrinking, that is, increasingly focused;
- Using the available algorithms—like factor analysis in SPSS and core-analysis in Pajek—it was not possible to delineate nano-journals clearly from other journals relevant in the direct environment, such as disciplinary journals in chemistry and physics, and general science journals like *Science* and *Nature.* The nano journals exhibit factorial complexity to a larger extent than disciplinary journals, and general science journals are interwoven with the interdisciplinary interface between the relevant disciplines (physics and chemistry). The Library of Congress classification does not yet follow the new developments of nano-journals except when these journals have the stem “nano” in their title.
- Using betweenness centrality in the citing-dimension, we found a set of ten journals positioned together in the vector space at the interface between physics and chemistry, and delineated from general science journals. We analyzed the national contributions to this set and found the U.S.A. to be the leading nation in nanotechnology, while Japan has lost its second place to China. However, other nations (e.g., Taiwan and South Korea) have also been able to increase their participation in this scientific literature.

[9] The percentage of non-US inventors is 40.4%; for the much smaller set of assignees this percentage is 40.1%.

We hope with the above not only to have provided new insights into the development of nanoscience and nanotechnology as a field of science at the interface between physics and chemistry (Zitt & Bassecoulard, 2006; Porter *et al*., 2006), but also into how to delineate the interdisciplinary journals at this interface from the multidisciplinary ones in general science (Leydesdorff, 2006a). This methodological contribution, of course, needs to be validated in other fields of science (Goldstone & Leydesdorff, forthcoming) and for additional years, for example, when the 2005 data of the Journal Citation Reports becomes available.

**Acknowledgment**

The authors are grateful to Stephen Bensman for supplying the information contained in Table 5, and to Martin Meyer for comments on an earlier draft.